\documentclass[runningheads]{miccai_llncs}

\usepackage{graphicx}
\usepackage[toc]{glossaries}
\usepackage[hidelinks]{hyperref}
\hypersetup{
    colorlinks=true,
    linkcolor=blue,
    filecolor=magenta,      
    urlcolor=cyan,
}

\usepackage{amsmath}
\usepackage{amssymb}
\usepackage{bbm}
\usepackage{eucal}
\usepackage{enumitem}
\usepackage{booktabs}
\usepackage{caption,subcaption}
\usepackage{bbm}
\usepackage{tikz}
\usepackage{lipsum}
\usepackage{adjustbox}

\DeclareMathAlphabet\mathbfcal{OMS}{cmsy}{b}{n}

\usepackage{tcolorbox}
\usepackage{nicefrac,xfrac}

\newcommand{\figref}[1]{Figure~\ref{#1}}
\newcommand{\tableref}[1]{Table~\ref{#1}}
\renewcommand{\eqref}[1]{Equation~\ref{#1}}
\newcommand{\secref}[1]{Section~\ref{#1}}

\newcommand{\mycomment}[1]{}

\usepackage{array}
\usepackage{multirow}
\newcolumntype{L}[1]{>{\raggedright\let\newline\\\arraybackslash\hspace{0pt}}m{#1}}
\newcolumntype{C}[1]{>{\centering\let\newline\\\arraybackslash\hspace{0pt}}m{#1}}
\newcolumntype{R}[1]{>{\raggedleft\let\newline\\\arraybackslash\hspace{0pt}}m{#1}}

\usepackage{array}
\usepackage{multirow}
\newcolumntype{L}[1]{>{\raggedright\let\newline\\\arraybackslash\hspace{0pt}}m{#1}}
\newcolumntype{C}[1]{>{\centering\let\newline\\\arraybackslash\hspace{0pt}}m{#1}}
\newcolumntype{R}[1]{>{\raggedleft\let\newline\\\arraybackslash\hspace{0pt}}m{#1}}

\makeglossaries

\newacronym[plural=CNNs,firstplural=Convolutional Neural Networks (CNNs)]{cnn}{CNN}{Convolutional Neural Network}

\newacronym[plural=FCNs,firstplural=Fully Convolutional Networks (FCNs)]{fcn}{FCN}{Fully Convolutional Network}

\newacronym{dl}{DL}{deep learning}
\newacronym{qcmlb}{QC-MLB}{Question-Centric Multimodal Low-rank Bilinear}
\newacronym{bert}{BERT}{Bidirectional Encoder Representations from Transformers}
\newacronym{bleu}{BLEU}{Bilingual Evaluation Understudy}
\newacronym{mlm}{MLM}{Masked Language Model}
\newacronym{nsp}{NSP}{Next Sentence Prediction}
\newacronym{relu}{ReLU}{rectified linear unit}
\newacronym{lrelu}{LeakyReLU}{leaky rectified linear unit}
\newacronym{nn}{NN}{neural network}
\newacronym{chal}{ImageCLEF-VQA-Med}{ImageCLEF-VQA-Med}
\newacronym{proposed}{\textit{\textless~Model~\textgreater}}{\textbf{Full name of the proposed model}}
\newacronym{mri}{MRI}{magnetic resonance imaging}
\newacronym{kits}{KiTS19}{Kidney Tumor Segmentation Challenge 2019}
\newacronym{ibsr}{IBSR18}{Internet Brain Segmentation Repository}
\newacronym{hene}{HeNe}{Ume{\aa} Head and Neck}
\newacronym{pros}{ProST}{Ume{\aa} Prostate}

\newacronym{ct}{CT}{computed tomography}
\newacronym{t1c}{T1c}{post-contrast T1-weighted}
\newacronym{t2}{T2w}{T2-weighted}
\newacronym{t1}{T1w}{T1-weighted}
\newacronym{flair}{FLAIR}{T2  Fluid  Attenuated  Inversion  Recovery}
\newacronym{lgg}{LGG}{low-grade glioma}
\newacronym{hgg}{HGG}{high-grade glioma}

\newacronym{svm}{SVMs}{Support-vector Machines}
\newacronym{crf}{CRF}{Conditional Random Field}
\newacronym{vae}{VAE}{Variational Auto-Encoder}
\newacronym{tunet}{TuNet}{End-to-end Hierarchical Tumor Segmentation using Cascaded Networks}
\newacronym{dram}{DRAM}{dynamic random access memory}
\newacronym[plural=GPUs,firstplural=graphics processing units (GPUs)]{gpu}{GPU}{graphics processing unit}

\newacronym{hpc2n}{HPC2N}{High Performance Computer Center North}

\newacronym{snic}{SNIC}{Swedish National Infrastructure for Computing}

\newacronym{dsc}{DSC}{S{\o}rensen-Dice coefficient}
\newacronym{se}{SE}{standard error}
\newacronym{seb}{SEB}{Squeeze-and-Excitation block}
\newacronym[plural=REBs,firstplural=ResNet blocks]{res}{REB}{ResNet block}
\newacronym{hd}{HD}{Hausdorff distance}
\newacronym{hd95}{HD95}{$95^{th}$ percentile of the Hausdorff distance}
\newacronym{dauc}{DAUC}{Dice Area Under Curve}
\newacronym{rftp}{RFTPs}{Ratio of Filtered True Positives}
\newacronym{rftn}{RFTNs}{Ratio of Filtered True Negatives}

\newacronym[plural=SDs,firstplural=standard deviations (SDs)]{sd}{SD}{standard deviation}
\newacronym{brats}{BraTS 2020}{Brain Tumors in Multimodal Magnetic Resonance Imaging Challenge 2020}

\newacronym{mdnet}{MDNet}{End-to-end Multi-Decoder Cascaded Network for Tumor Segmentation}
\newacronym{ce}{CE}{categorical cross--entropy}

\newacronym{miccai}{MICCAI}{International Conference on Medical Image Computing and Computer Assisted Intervention}
\usepackage{scalefnt}

\begin{document}

\title{Multi-Decoder Networks with Multi-Denoising Inputs for Tumor Segmentation}

\author{Minh H. Vu\inst{1} \and Tufve Nyholm\inst{1} \and Tommy L\"{o}fstedt\inst{2}}

\institute{Department of Radiation Sciences, Ume{\aa} University, Ume{\aa}, Sweden \and
Department of Computing Science, Ume{\aa} University, Ume{\aa}, Sweden \\
\email{tommy@cs.umu.se}}
\maketitle

\begin{abstract}

Automatic segmentation of brain glioma from multimodal MRI scans plays a key role in clinical trials and practice. Unfortunately, manual segmentation is very challenging, time-consuming, costly, and often inaccurate despite human expertise due to the high variance and high uncertainty in the human annotations. In the present work, we develop an end-to-end deep-learning-based segmentation method using a multi-decoder architecture by jointly learning three separate sub-problems using a partly shared encoder. We also propose to apply smoothing methods to the input images to generate denoised versions as additional inputs to the network. The validation performance indicate an improvement when using the proposed method. The proposed method was ranked 2nd in the task of Quantification of Uncertainty in Segmentation in the \glsdesc{brats}.

\end{abstract}



\section{Introduction}
\label{sec:intro}

Glioma is a particular kind of brain tumor that develops from glial cells. It is the most frequently occurring type of brain tumor and the one with the highest mortality rate. Glioma are categorized by the World Health Organization (WHO) into four grades: \gls{lgg} (class I and II), and \gls{hgg} (class III and IV), where \gls{hgg} is being considered a dangerous and life-threatening tumor. Specifically, about 190,000 cases occur annually worldwide~\cite{Castells2009AutomatedBT}, and around 90~\%~\cite{Thurnher2012} of patients die within 24 months of surgical resection. Segmentation of the tumor plays a role both for radiotherapy treatment planning and for diagnostic follow-up of the disease. Manual segmentation is time-consuming, subjective, and associated with uncertainties due to the variation of shape, location, and appearance of the tumors. Hence, decision support or automating the segmentations may improve the treatment quality as well as enhancing the efficiency when handling this patient group.

Inspired by a need of automatic segmentation of brain tumors in multimodal \gls{mri} scans, the \gls{brats}~\cite{menze2014multimodal,bakas2017advancing,bakas2018identifying,bakas2017segmentation,bakas2017segmentation17}  is a yearly challenge
(associated with the \gls{miccai}) that aims to evaluate state-of-the-art methods for brain tumor segmentation. \gls{brats} provides the participants with images from four structural \gls{mri} modalities: \gls{t1c}, \gls{t2}, \gls{t1}, and \gls{flair} for brain tumor analysis and segmentation. Masks were annotated manually by one to four raters followed by improvements by expert raters. The segmentation performances of the participants were evaluated using the \gls{dsc}, sensitivity, specificity, and the \gls{hd95}.

\begin{figure}[!t]
    \begin{center}
        \includegraphics[width=\textwidth]{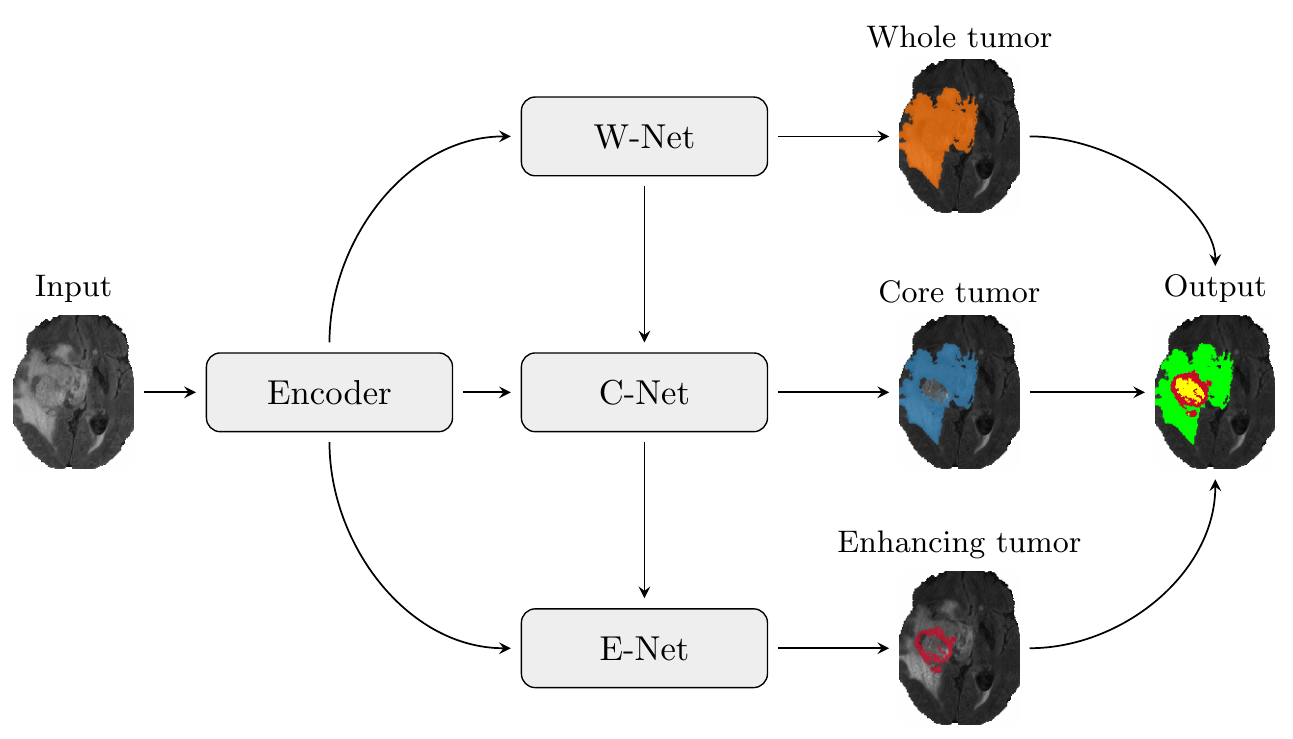}
    \end{center}
    \caption[]{Schematic visualization of the \acrshort{mdnet} architecture.}
    \label{fig:method}
\end{figure}

Since the introduction of the U-Net by Ronneberger \textit{et al.}~\cite{ronneberger2015unet}, \glspl{cnn} incorporating skip connections have become the baseline architecture for medical image segmentation. Various architectures, often building on or extending this baseline, have been proposed to address the brain tumor segmentation problem. In BraTS 2019, Jiang \textit{et al.}~\cite{jiang2020brats19}, who was the first-place winner of the challenge, proposed an end-to-end two-stage cascaded U-Net to segment the substructures of brain tumors from coarse (in the first stage) to fine (in the second stage) prediction. In the same challenge, Zhao \textit{et al.}~\cite{zhao2020brats19}, who won the second place, introduced numerous tricks for 3D \gls{mri} brain tumor segmentation including processing methods, model designing methods, and optimizing methods. McKinley \textit{et al.}~\cite{mckinley2020brats19} proposed DeepSCAN, which is a modification of their previous 3D-to-2D \gls{fcn}, by replacing batch normalization with instance normalization and adding a lightweight local attention mechanism to secure the third place in the BraTS 2019.

The architecture proposed in this work is an extension of the one in~\cite{vu2020brats19} from the BraTS 2019 where \gls{tunet} was introduced. Despite achieving a decent performance, the main drawback of \gls{tunet} is that it comprises three cascaded networks that make it hard to fit a full volume, with shape $240 \times 240 \times 155$, into memory on any recent \glspl{gpu}. Because of this, the \gls{tunet} adapted a patch-based segmentation approach, leading to long training times. In addition to that, the \gls{tunet} might suffer from a lack of global information about the image.

Motivated by the successes of the cascaded networks, presented in \textit{e.g.}~\cite{jiang2020brats19,vu2020brats19}, the present work proposes a multi-decoder architecture, denoted \gls{mdnet}, to separate a complicated problem into simple sub-problems. We also propose to use multiple denoised versions of the original images as inputs to the network. The hypothesis was that this would counteract the salt and pepper noise often seen in \gls{mri} scans~\cite{ali2016remove}. To the best of our knowledge, this is the first use of this technique.

The authors hypothesize that the \gls{mdnet} will reduce overfitting problems by employing a shared encoder between three different decoders, while denoised \gls{mri} images will help the network to gain more insight into the multimodal input images with the presence of another two versions of the images: (i) a salt and pepper-free one from the use of a median filter, and (ii) one with reduced high-frequency components by employing a low-pass Gaussian filter.


\section{Methods}
\label{sec:method}

Inspired by the drawbacks of the method proposed in~\cite{vu2020brats19}, the authors here also propose an end-to-end framework that separates the complicated multi-class tumor segmentation problem into three simpler binary segmentation problems, but with major change in the design. The \gls{mdnet} consumes much less memory compared to the \gls{tunet}, which means that whole input volumes can be fit into the \gls{gpu} memory. Hence, the proposed \gls{mdnet} is able to take advantage of global details. In addition to that, the design of \gls{mdnet} results in shorter training times since it uses whole volumes instead of patches, as was the case with the \gls{tunet}.


\subsection{Encoder Network}
\label{subsec:framework}

The encoder network consists of conventional convolution blocks~\cite{ronneberger2015unet}, where each block includes a convolution layer with batch normalization and a \gls{lrelu} activation function. Each convolutional block is then followed by a \gls{seb} block (see \secref{subsec:se}). Max-pooling layers were used for downsampling. All convolutional filters had the size of $3\times3\times3$, and the initial numbers of filters were set to twelve, which in the proposed architecture is equivalent to three denoising methods applied to the four given modalities (see \secref{subsec:denoising}). The encoder output has shape $96\times20\times24\times16$. The complete architecture of the proposed encoder network is detailed in \tableref{tab:encoder}.

\begin{table}[!th]
    \caption{The encoder architecture. ``Conv3'' denotes a $3\times3\times3$ convolution, ``BN'' stands for batch normalization, ``LeakyReLU'' is the \glsdesc{lrelu}, and ``SEB'' denotes the \glsdesc{seb} (see \secref{subsec:se}).}
    \centering
    \begin{tabular}{l C{0.25cm} C{4.8cm} c C{3.5cm}}
        \toprule
        Name      && Layers 				  & Repeat & Output size \\
        \cmidrule{1-1}  \cmidrule{3-5}
        Input     &&    					  &        & $12 \times 160 \times 192 \times 128$ \\
        EncBlk--0 && Conv3, BN, LeakyReLU, SEB & 2      & $12 \times 160 \times 192 \times 128$ \\
        EncDwn--1 && MaxPooling       		  & 1      & $12 \times 80 \times 96 \times 64$ \\
        EncBlk--1 && Conv3, BN, LeakyReLU, SEB & 2      & $24 \times 80 \times 96 \times 64$ \\
        EncDwn--2 && MaxPooling       		  & 1      & $24 \times 40 \times 48 \times 32$ \\
        EncBlk--2 && Conv3, BN, LeakyReLU, SEB & 2      & $48 \times 40 \times 48 \times 32$ \\
        EncDwn--3 && MaxPooling       		  & 1      & $48 \times 20 \times 24 \times 16$ \\
        EncBlk--3 && Conv3, BN, LeakyReLU, SEB & 2      & $96 \times 20 \times 24 \times 16$ \\ 
        \bottomrule
    \end{tabular}
    \label{tab:encoder}
\end{table}

\begin{table}[!th]
    \centering
    \caption{Decoder architectures. Here, ``Conv3'' means a $3\times3\times3$ convolution, ``Conv1'' a $1\times1\times1$ convolution, ``BN'' denotes for batch normalization, ``LeakyReLU'' means the \glsdesc{lrelu}, ``SEB'' denotes the \glsdesc{seb} (see \secref{subsec:se}), ``Up--\{$X$\}'' represents the 3D linear spatial upsampling of block $X$, $(+)$ denotes the concatenation operation. In the name column, $W$--, $C$-- and $E$-- correspond to the whole, core, and enhancing tumor regions, respectively.
    }
    \begin{tabular}{l C{0.25cm} C{4.8cm} c C{3.5cm}}
    \toprule
    Name 			 && Layers 					    & Repeat & Output size\\ 
    \cmidrule{1-1}      \cmidrule{3-5}
    W--DecCat--2     && Up--EncBlk--3 + EncBlk--2   & 1      & $144 \times 40 \times 48 \times 32$ \\
    W--DecSae--2     && SEB      				    & 1      & $144 \times 40 \times 48 \times 32$ \\
    W--DecBlk--2     && Conv3, BN, LeakyReLU, SEB    & 2      & $48 \times 40 \times 48 \times 32$ \\
    W--DecCat--1     && Up--DecBlk--2 + EncBlk--1   & 1      & $72 \times 80 \times 96 \times 64$ \\
    W--DecSae--1     && SEB      				    & 1      & $72 \times 80 \times 96 \times 64$ \\
    W--DecBlk--1     && Conv3, BN, LeakyReLU, SEB    & 2      & $24 \times 80 \times 96 \times 64$ \\
    W--DecCat--0     && Up--DecBlk--1 + EncBlk--0   & 1      & $36 \times 160 \times 192 \times 128$ \\
    W--DecSae--0     && SEB      				    & 1      & $36 \times 160 \times 192 \times 128$ \\
    W--DecBlk--0     && Conv3, BN, LeakyReLU, SEB    & 2      & $12 \times 160 \times 192 \times 128$ \\
    W--Output        && Conv1, Sigmoid       	    & 1      & $1 \times 160 \times 192 \times 128$ \\
    \cmidrule{1-1}      \cmidrule{3-5}
    C--DecCat--2     && W--DecBlk--2 + W--DecCat--2 & 1      & $192 \times 40 \times 48 \times 32$ \\
    C--DecSae--2     && SEB      					& 1      & $192 \times 40 \times 48 \times 32$ \\
    C--DecBlk--2     && Conv3, BN, LeakyReLU, SEB    & 2      & $48 \times 40 \times 48 \times 32$ \\
    C--DecCat--1     && W--DecBlk--1 + W--DecCat--1 & 1      & $96 \times 80 \times 96 \times 64$ \\
    C--DecSae--1     && SEB      					& 1      & $96 \times 80 \times 96 \times 64$ \\
    C--DecBlk--1     && Conv3, BN, LeakyReLU, SEB    & 2      & $24 \times 80 \times 96 \times 64$ \\
    C--DecCat--0     && W--DecBlk--0 + W--DecCat--0 & 1      & $48 \times 160 \times 192 \times 128$ \\
    C--DecSae--0     && SEB                          & 1      & $48 \times 160 \times 192 \times 128$ \\
    C--DecBlk--0     && Conv3, BN, LeakyReLU, SEB    & 2      & $12 \times 160 \times 192 \times 128$ \\
    C--Output        && Conv1, Sigmoid              & 1      & $1 \times 160 \times 192 \times 128$ \\
    \cmidrule{1-1}      \cmidrule{3-5}
    E--DecCat--2     && C--DecBlk--2 + W--DecCat--2 & 1      & $240 \times 40 \times 48 \times 32$ \\
    E--DecSae--2     && SEB                          & 1      & $240 \times 40 \times 48 \times 32$ \\
    E--DecBlk--2     && Conv3, BN, LeakyReLU, SEB    & 2      & $48 \times 40 \times 48 \times 32$ \\
    E--DecCat--1     && C--DecBlk--1 + W--DecCat--1 & 1      & $96 \times 80 \times 96 \times 64$ \\
    E--DecSae--1     && SEB                          & 1      & $96 \times 80 \times 96 \times 64$ \\
    E--DecBlk--1     && Conv3, BN, LeakyReLU, SEB    & 2      & $24 \times 80 \times 96 \times 64$ \\
    E--DecCat--0     && C--DecBlk--0 + W--DecCat--0 & 1      & $48 \times 160 \times 192 \times 128$ \\
    E--DecSae--0     && SEB                          & 1      & $48 \times 160 \times 192 \times 128$ \\
    E--DecBlk--0     && Conv3, BN, LeakyReLU, SEB    & 2      & $12 \times 160 \times 192 \times 128$ \\
    E--Output        && Conv1, Sigmoid              & 1      & $1 \times 160 \times 192 \times 128$ \\
    \bottomrule
    \end{tabular}
    \label{tab:e_decoder}
\end{table}


\subsection{Multi-decoder Networks}
\label{subsec:decoder_networks}

\tableref{tab:e_decoder} illustrates the proposed multi-decoder networks. The decoder networks include three separate paths, where each path is employed to cope with a specific aforementioned tumor region including whole, core, and enhancing, that are denoted by W-Net, C-Net, and E-Net, respectively. Each decoder path comprises skip connections as in U-Net. There was also a \gls{seb} after each convolution block and a concatenation operation of the output of the spatial upsampling layers with the feature maps from the encoder at the same level. To enrich the feature maps at the beginning of each level in the C-Net, the feature map at the end of the W-Net on the same level is used. A similar approach is employed in the decoder network of the E-Net and C-Net. By utilizing these, we hypothesize that the W-Net will constrain the C-Net, while the C-Net will constrain the E-Net.
\figref{fig:method} illustrates the proposed architecture.


\subsection{Squeeze-and-Excitation Block}
\label{subsec:se}

We added a \gls{seb} as proposed by Hu \textit{et al.}~\cite{hu2018squeeze} after each convolution block or concatenation operation. The idea of \gls{seb} is to adapt the weight of each channel in a feature map by adding a content-aware mechanism at almost no computational cost. In recent days, \gls{seb} has been widely employed to achieve a huge boost in performance. A conventional \gls{seb} includes the following layers in sequence: global pooling, fully connected, \gls{relu} activation function, fully connected, and a sigmoid activation function~\cite{hu2018squeeze}.


\subsection{Denoising the Inputs}
\label{subsec:denoising}

The inputs to the network were the \gls{mri} modalities, and also each modality after denoising using two different methods: median denoising and Gaussian smoothing. The authors then concatenated the three versions of the images for each modality (the raw image, and the two denoised versions) to obtain a total of twelve images, that were input as different channels. For the median denoising, we used a $3\times3\times3$ median filter; the Gaussian smoothing used a $3\times3\times3$ Gaussian filter with a standard deviation of 0.5. In this sense, adding a Gaussian smoothed version of the input is similar to adding a down-scaled version of the input image as was proposed for the \gls{tunet}~\cite{vu2020brats19}.


\subsection{Preprocessing and Augmentation}
\label{subsec:prep}

All input images were normalized to have a mean zero and unit variance. In order to reduce overfitting and increase the diversity of data available for training models, we used on-the-fly data augmentation~\cite{isensee_fabian_2020_3632567} comprising: (1) randomly rotating the images in the range $[-1, 1]$ degrees on all three axes, (2) random mirror flipping with a probability of 0.5 on all three axes, (3) elastic transformation with a probability of 0.3, (4) random scaling in the range $[0.9, 1.1]$ with a probability of 0.3, and (5) random cropping with subsequent resizing with a probability of 0.3.

As in~\cite{Simard2003elastic}, the elastic transformations used a random displacement field, $\Delta$, such that
\begin{equation}
    R_w = R_o + \alpha \Delta,
\end{equation}
where $\alpha$ is the strength of the displacement, while $R_w$ and $R_o$ denote the location of a voxel in the warped and original image, respectively. For each axis, a random number was drawn uniformly in $[-1, 1]$ such that $\Delta_x \sim \mathcal{U}(-1, 1)$, $\Delta_y \sim \mathcal{U}(-1, 1)$, and $\Delta_z \sim \mathcal{U}(-1, 1)$. The displacement field was finally convolved with a Gaussian kernel having standard deviation $\sigma$. In the present case, $\alpha=1$ and $\sigma=0.25$.


\subsection{Post-processing}
\label{subsec:post}

The most challenging task of \gls{brats} specifically, and BraTS challenges in general, is to distinguish between \gls{lgg} and \gls{hgg} patients by labeling small vessels lying in  the tumor core as edema or necrosis. In order to tackle this problem, we used the same strategy as proposed in our previous work~\cite{vu2020brats19}. In specific, we labeled all small enhancing tumor region with less than 500 connected voxels as necrosis. The proposed post-processing step aims to handle a few cases where the proposed networks fail to differentiate between the whole and core tumor regions.

\subsection{Task 3: Quantification of Uncertainty in Segmentation}
\label{subsec:task3}

The organizers of the BraTS challenge introduced the task of ``Quantification of Uncertainty in Segmentation'' in BraTS 2019 and was held again in BraTS 2020. This task is aimed to measure the uncertainty in the context of glioma region segmentation by rewarding predictions that are (a) confident when correct and (b) uncertain when incorrect. Participants were expected to generate uncertainty maps in the range of $[0, 100]$, where 0 represents the most certain and 100 represents the most uncertain. The performance was evaluated based on three metrics: \gls{dauc}, \gls{rftp}, and \gls{rftn}.

Similar to~\cite{vu2020brats19}, the proposed network, \gls{mdnet}, predicts the probability of three tumor regions, it thus benefits from this task. Following~\cite{vu2020brats19}, an uncertainty score, $u^r_{i,j,k}$, at voxel $(i,j,k)$ is defined by
\begin{equation}
     u^r_{i,j,k} =   
     \begin{cases} 
         200 (1-p^{r}_{i,j,k}), & \text{if } p^{r}_{i,j,k} \geq 0.5, \\
         200 p^{r}_{i,j,k},     & \text{if } p^{r}_{i,j,k} < 0.5,
   \end{cases}
\end{equation}
where $u^r_{i,j,k} \in [0, 100]^{|\mathcal{R}|}$ and $p^{r}_{i,j,k} \in [0, 1]^{|\mathcal{R}|}$ are the uncertainty score map and probability map, respectively. Here, $r \in \mathcal{R}$, where $\mathcal{R}$ is the set of tumor regions, \textit{i.e.} whole, core, and enhancing region.


\section{Experiments}
\label{subsec:exp}



\subsection{Implementation Details and Training}
\label{subsec:train}

The proposed method was implemented in Keras 2.2.4\footnote{\url{https://keras.io}} with TensorFlow 1.12.0\footnote{\url{https://tensorflow.org}} as the backend. The experiments were trained on NVIDIA Tesla V100 \glspl{gpu} from the \gls{hpc2n} at Ume{\aa} University, Sweden. Seven models were trained from scratch for $N_e=200$ epochs, with a mini-batch size of one. The training time for a single model was about six days.


\subsection{Loss}
\label{subsec:loss}

For evaluation of the segmentation performance, we used a combination of the \gls{dsc} loss and \gls{ce} as the loss function. The \gls{dsc} is defined as~\cite{vu2019evaluation,vu2020brats19}
\begin{equation}\label{eq:dice}
    D(u, v)=\frac{2 \cdot |u \cap v|}{|u| + |v|},
\end{equation}
where $u$ and $v$ are the output segmentation and its corresponding ground truth, respectively. To include the the \gls{dsc} in the loss function, we employed the soft \gls{dsc} loss, which is defined as~\cite{isensee2019nonew,vu2019evaluation,vu2020brats19}
\begin{equation} \label{eqn:dscloss}
    \mathcal{L}_{DSC}(u, v) = \frac{-2 \sum_i u_i v_i}{\sum_i u_i + \sum_i v_i + \epsilon},
\end{equation}
where for each label $i$, the $u_i$ is the softmax output of the proposed network for label $i$, $v$ is a one-hot encoding of the ground truth labels (segmentation maps in this case), and $\epsilon = 1 \cdot 10^{-5}$ is a small constant added to avoid division by zero.

Following~\cite{isensee2019nonew,vu2019evaluation}, for unbalanced data sets with small structures like in the \gls{brats} data, we added the \gls{ce} term to our loss function to make the loss surface smoother. The \gls{ce} is defined as
\begin{equation} \label{eqn:celoss}
    \mathcal{L}_{CE}(u, v) = -\sum_i u_i \cdot \text{log} (v_i).
\end{equation}
The combination of the \gls{dsc} loss and \gls{ce} (denoted a \textit{hybrid loss}) is simply defined as the sum of the two losses, as
\begin{equation} \label{eqn:hybrid_loss}
    \mathcal{L}_{\mathrm{hybrid}}(u, v) = \mathcal{L}_{DSC}(u, v) + \mathcal{L}_{CE}(u, v).
\end{equation}
The final loss function that was used for training contained one hybrid loss for each tumor region, and was thus
\begin{equation} \label{eqn:hybridloss}
    \mathcal{L}(u, v) = \sum_{r\in\mathcal{R}} \mathcal{L}_{\mathrm{hybrid}}(u_r, v_r),
\end{equation}
where $\mathcal{R}$ again is the set of tumor regions (the whole, core, and enhancing regions) and $\mathcal{L}_{\mathrm{hybrid}}(u_r, v_r)$ is the hybrid loss for a particular tumor region.

The segmentation performance was also evaluated using the \gls{hd95}, a common metric for evaluating segmentation performances. The \gls{hd} is defined as~\cite{Hausdorff2008}
\begin{equation}
    H(u,v) = \max \{d(u, v), d(v, u)\},
\end{equation}
where 
\begin{equation}
    d(u, v) = \max_{u_i \in u} {\min_{v_i \in v} \|u_i - v_i\|_2},
\end{equation}
in which $\|u_i - v_i\|_2$ is the spatial Euclidean distance between points $u_i$ and $v_i$ on the boundaries of output segmentation $u$ and ground truth $v$.


\subsection{Optimization}
\label{subsec:optim}

The authors used the Adam optimizer~\cite{kingma2014adam} with an initial learning rate of $\alpha_0  = 1 \cdot 10^{-4}$ and momentum parameters of $\beta_1=0.9$ and $\beta_2=0.999$. Following Myronenko \textit{et al.} in~\cite{Myronenko2018}, the learning rate was
decayed as
\begin{equation}
    \alpha_e = \alpha_0 \cdot \left( 1 - \frac{e}{N_e} \right)^3,
\end{equation}
where $e$ and $N_e=200$ are epoch counter and total number of epochs, respectively. 

The authors also used $L_2$ regularization with a penalty parameter of $1 \cdot 10^{-5}$, which was applied to the kernel weight matrices, for all convolutional layers to counter overfitting. The activation function of the final layer was the logistic sigmoid function.


\section{Results and Discussion}
\label{sec:results}

\tableref{tab:cv} shows the mean \gls{dsc} and \gls{hd95} scores and \glspl{sd} computed from the five-folds of cross-validation on 396 cases of the training set. From \tableref{tab:cv} we see that: (i) the U-Net with denoised input improved the \gls{dsc} and \gls{hd95} on all tumor regions, and (ii) the proposed model with denoising boosted the performance in both metrics (\gls{dsc} and \gls{hd95}) by a large margin.

\begin{table}[!th]
    \def\width{1 cm}
    \def\widthdetail{3.7 cm}
    \caption{Mean \gls{dsc} (higher is better) and \gls{hd95} (lower is better) and their \glspl{se} (in parentheses) computed from the five-folds of cross-validation on the training set (369 cases) for the different models.
    }
    \centering
    \begin{tabular}{L{\widthdetail} C{1.0em} C{\width} C{\width} C{\width} C{1.0em} C{\width} C{\width} C{\width}}
    \toprule
          && \multicolumn{3}{c}{\gls{dsc}} && \multicolumn{3}{c}{\gls{hd95}} \\
    Model && whole  		  & core   			 & enh.   			&& whole  	 	& core   	   & enh.   \\
    \cmidrule{1-1}\cmidrule{3-5}\cmidrule{7-9}
    U-Net without denoising                     && 90.66 (0.38)   & 86.93 (0.71)     & 76.16 (1.37)     && 4.91 (0.41)  & 4.78 (0.42)  & 3.46 (0.31)    \\
    U-Net with denoising                     	&& 90.98 (0.31)   & 87.53 (0.68)     & 76.55 (1.36)     && 4.49 (0.26)  & 4.32 (0.29)  & 3.41 (0.29)    \\
    Proposed with denoising                     && 92.75 (0.25)   & 88.34 (0.70)     & 78.13 (1.32)     && 4.32 (0.29)  & 4.30 (0.31)  & 3.29 (0.24)    \\
    \bottomrule
    \end{tabular}
    \label{tab:cv}
\end{table}

\tableref{tab:valid} shows the mean \gls{dsc} and \gls{hd95} scores on the validation set, computed on the predicted masks by the evaluation server\footnote{\url{https://www.cbica.upenn.edu/BraTS20/lboardValidation.html}} (team name \textit{UmU}). The BraTS 2020 final validation dataset results were 90.55, 82.67 and 77.17 for the average \gls{dsc}, and 4.99, 8.63 and 27.04 for the average \gls{hd95}, for whole tumor, tumor core and enhanced tumor core, respectively. These results were slightly lower than the top-ranking teams.

\tableref{tab:valid_unc} provides the mean \gls{dauc}, \gls{rftp}, and \gls{rftn} scores on the validation set obtained after uploading the predicted masks and corresponding uncertainty maps to the evaluation server\footnote{\url{https://www.cbica.upenn.edu/BraTS20/lboardValidationUncertainty.html}}. As can be seen from \tableref{tab:valid_unc}, the \gls{rftn} scores were the best amongst the best-ranking participants.

\tableref{tab:test} and \tableref{tab:test_unc} show the mean \gls{dsc} and \gls{hd95}, and the mean \gls{dauc}, \gls{rftp}, and \gls{rftn} scores on the test set, respectively. In the task of Quantification of Uncertainty in Segmentation, our proposed method was ranked 2nd.

\begin{table}[!th]
\def\width{1. cm}
\def\widthdetail{3.7 cm}
\caption{Results of Segmentation Task on \gls{brats} validation data (125 cases). The results were obtained by computing the mean of predictions of seven models trained from the scratch. ``UmU'' denotes the name of our team.  The metrics were computed by the online evaluation platform. All the predictions were post-processed before submitting to the server. The top rows correspond to the top-ranking teams from the online system retrieved at 11:38:02 EDT on August 3, 2020.}
\centering
    \begin{tabular}{l C{1.0em} C{\width} C{\width} C{\width} C{1.0em} C{\width} C{\width} C{\width}}
    \toprule
                                                && \multicolumn{3}{c}{\gls{dsc}} && \multicolumn{3}{c}{\gls{hd95}} \\
    Team                                        && whole   & core  & enh.     && whole     & core   & enh.   \\
    \cmidrule{1-1}\cmidrule{3-5}\cmidrule{7-9}
    deepX                                       &  & 91.02 & 85.00 & 78.53    && 4.44	   & 5.90   & 24.06  \\
    Radicals                                    &  & 90.82 & 84.96 & 78.69	  && 4.71      & 8.56   & 35.01	 \\
    WassersteinDice	                            &  & 90.58 & 83.79 & 78.01	  && 4.74      & 8.96   & 27.02	 \\
    CKM	                                        &  & 90.83 & 83.82 & 78.59	  && 4.87      & 5.97   & 26.57  \\
    \cmidrule{1-1}\cmidrule{3-5}\cmidrule{7-9}
    UmU                                         &  & 90.55 & 82.67 & 77.17    && 4.99      & 8.63   & 27.04  \\
    \bottomrule
\end{tabular}
\label{tab:valid}
\end{table}

\begin{table}[!th]
\def\width{1. cm}
\def\widthdetail{1.7 cm}
\caption{Results of Quantification of Uncertainty Task on \gls{brats} validation data (125 cases) including mean \gls{dauc} (higher is better), \gls{rftp} (lower is better) and \gls{rftn} (lower is better). The results were obtained by computing the mean of predictions of seven models trained from scratch. ``UmU'' denotes the name of our team and the ensemble of seven models.  The metrics were computed by the online evaluation platform. The top rows correspond to the top-ranking teams from the online system retrieved at 11:38:02 EDT on August 3, 2020.}
\centering
    \begin{tabular}{l C{0.725em} C{\width} C{\width} C{\width} C{0.25em} C{\width} C{\width} C{\width} C{0.25em} C{\width} C{\width} C{\width}}
    \toprule
                                                && \multicolumn{3}{c}{DAUC}   && \multicolumn{3}{c}{\gls{rftp}}  && \multicolumn{3}{c}{\gls{rftn}} \\
    Team                                        && whole    & core   & enh.   && whole  & core   & enh.          && whole  & core   & enh.      \\
    \cmidrule{1-1}\cmidrule{3-5}\cmidrule{7-9}\cmidrule{11-13}
    med\_vision                                 &  & 95.24  & 92.23  & 83.24  && 0.28   & 0.62     & 0.93        && 87.74  & 98.74  & 98.74      \\
    nsu\_btr                                    &  & 93.58  & 90.04  & 85.14  && 35.72  & 48.18    & 9.59        && 98.44  & 98.60  & 98.64      \\
    SCAN                                        &  & 93.46  & 82.98  & 80.64  && 12.40  & 19.95    & 21.53       && 0.87   & 0.42   & 0.24      \\
    \cmidrule{1-1}\cmidrule{3-5}\cmidrule{7-9}\cmidrule{11-13}
    UmU                                         &  & 92.59  & 83.61  & 78.83  && 4.48   & 10.13    & 7.95        && 0.27   & 0.17	& 0.08      \\
    \bottomrule
\end{tabular}
\label{tab:valid_unc}
\end{table}

\begin{table}[!th]
\def\width{1. cm}
\def\widthdetail{3.7 cm}
\caption{Results of Segmentation Task on \gls{brats} test data (166 cases). The results were obtained by computing the mean of predictions of seven models trained from the scratch. The metrics were computed by the online evaluation platform. All the predictions were post-processed before submitting to the server.}
\centering
    \begin{tabular}{l C{1.0em} C{\width} C{\width} C{\width} C{1.0em} C{\width} C{\width} C{\width}}
    \toprule
                                                && \multicolumn{3}{c}{\gls{dsc}} && \multicolumn{3}{c}{\gls{hd95}} \\
    Team                                        && whole   & core  & enh.     && whole     & core   & enh.   \\
    \cmidrule{1-1}\cmidrule{3-5}\cmidrule{7-9}
    UmU                                         && 88.26   & 82.49 & 80.84    && 6.30      & 22.27  & 20.06  \\
    \bottomrule
\end{tabular}
\label{tab:test}
\end{table}

\begin{table}[!th]
\def\width{1. cm}
\def\widthdetail{1.7 cm}
\caption{Results of Quantification of Uncertainty Task on \gls{brats} test data (166 cases) including mean \gls{dauc} (higher is better), \gls{rftp} (lower is better) and \gls{rftn} (lower is better). The results were obtained by computing the mean of predictions of seven models trained from scratch. The metrics were computed by the online evaluation platform. }
\centering
    \begin{tabular}{l C{0.725em} C{\width} C{\width} C{\width} C{0.25em} C{\width} C{\width} C{\width} C{0.25em} C{\width} C{\width} C{\width}}
    \toprule
                                                && \multicolumn{3}{c}{DAUC}   && \multicolumn{3}{c}{\gls{rftp}}  && \multicolumn{3}{c}{\gls{rftn}} \\
    Team                                        && whole    & core   & enh.   && whole  & core   & enh.          && whole  & core   & enh.      \\
    \cmidrule{1-1}\cmidrule{3-5}\cmidrule{7-9}\cmidrule{11-13}
    UmU                                         &  & 90.61   & 85.83 & 83.03  && 4.18   & 5.49   & 4.45          && 0.31   & 1.68	& 0.07      \\
    \bottomrule
\end{tabular}
\label{tab:test_unc}
\end{table}

\section{Conclusion}
\label{sec:conclusion}

In this work, we proposed a multi-decoder network for segmenting tumor substructures from multimodal brain \gls{mri} images by separating a complex problem into simpler sub-tasks. The proposed network adopted a U-Net-like structure with Squeeze-and-Excitation blocks after each convolution and concatenation operation. We also proposed to stack original images with their denoised versions to enrich the input and demonstrated that the performance was boosted in both \gls{dsc} and \gls{hd95} metrics by a large margin. The results on the test set indicated that: (i) the proposed method performed competitively in the task of Segmentation, with \gls{dsc} scores of 88.26/82.49/80.84 and \gls{hd95} scores of 6.30/22.27/20.06 for the whole tumor, tumor core, and enhancing tumor core, respectively, (ii) the proposed method was  top 2 performing ones in the task of Quantification of Uncertainty in Segmentation.

\section*{Acknowledgement}
\label{sec:ack}

The computations were performed on resources provided by the \gls{snic} at the \gls{hpc2n} in Ume{\aa}, Sweden. We are grateful for the financial support obtained from the Cancer Research Fund in Northern Sweden, Karin and Krister Olsson, Ume\aa{} University, The V\"{a}sterbotten regional county, and Vinnova, the Swedish innovation agency.


\bibliographystyle{splncs04}
\bibliography{bib}

\end{document}